\documentstyle[aps,epsf]{revtex}

\begin{document}
\title{
  \begin{flushright} \begin{small}
    hep-th/0012151
  \end{small} \end{flushright}
Rotational Perturbations in Neveu-Schwarz--Neveu-Schwarz String
Cosmology}
\author{Chiang-Mei Chen\footnote{E-mail: cmchen@phys.ntu.edu.tw}}
\address{Department of Physics, National Taiwan University,
         Taipei 106, Taiwan}
\author{T. Harko\footnote{E-mail: tcharko@hkusua.hku.hk}}
\address{Department of Physics, The University of Hong Kong,
         Pokfulam, Hong Kong}
\author{M. K. Mak\footnote{E-mail: mkmak@vtc.edu.hk}}
\address{Department of Physics, The Hong Kong University of Science
         and Technology, Clear Water Bay, Hong Kong}
\date{December 8, 2000}
\maketitle

\begin{abstract}
First order rotational perturbations of the flat
Friedmann-Robertson-Walker (FRW) metric are considered in the
framework of four dimensional Neveu-Schwarz--Neveu-Schwarz (NS-NS)
string cosmological models coupled with dilaton and axion fields.
The decay rate of rotation depends mainly upon the dilaton field
potential $U$. The equation for rotation imposes strong
limitations upon the functional form of $U$, restricting the
allowed potentials to two: the trivial case $U=0$ and a
generalized exponential type potential. In these two models the
metric rotation function can be obtained in an exact analytic
form in both Einstein and string frames. In the potential-free
case the decay of rotational perturbations is governed by an
arbitrary function of time while in the presence of a potential
the rotation tends rapidly to zero in both Einstein and string
frames.
\end{abstract}



\section{Introduction}
On an astronomical scale rotation is a basic property of cosmic
objects. The rotation of planets, stars and galaxies inspired
Gamow to suggest that the Universe is rotating and the angular
momentum of stars and galaxies could be a result of the cosmic
vorticity \cite{Ga46}. But even that observational evidences of
cosmological rotation have been reported
\cite{Bi82,Bi83,NoRa97,Ku97}, they are still subject of
controversy. From the analysis of microwave background anisotropy
Collins and Hawking \cite{CoHa73} and Barrow, Juszkiewicz and
Sonoda \cite{BaJuSo85} have found some very tight limits of the
cosmological vorticity, $T_{obs}>3\times 10^5\, T_H$, where
$T_{obs}$ is the actual rotation period of our Universe and
$T_H=(1\sim 2)\times 10^{10}$ years is the Hubble time. Therefore
our present day Universe is rotating very slowly, if at all.

From theoretical point of view in 1949 G\"odel \cite{Go49} gave
his famous example of a rotating cosmological solution to the
Einstein gravitational field equations. The G\"odel metric,
describing a dust Universe with energy density $\rho$ in the
presence of a negative cosmological constant $\Lambda$ is
\begin{equation}\label{0}
ds^2 = \frac1{2\omega^2} \left[ -(dt + e^x dz)^2 + dx^2 + dy^2
     + \frac12 e^{2x} dz^2 \right].
\end{equation}
In this model the angular velocity of the cosmic rotation is
given by $\omega^2=4\pi\rho=-\Lambda$.  G\"odel also discussed
the possibility of a cosmic explanation of the galactic rotation
\cite{Go49}.  This rotating solution has attracted considerable
interest because the corresponding Universes possess the property
of closed timelike curves.

The investigation of rotating and rotating-expanding Universes
generated a large amount of literature in the field of general
relativity, the combination of rotation with expansion in
realistic cosmological models being one of the most difficult
tasks in cosmology (see \cite{Ob00} for a recent review of the
expansion-rotation problem in general relativity).  Hence
rotating solutions of the gravitational field equations cannot be
excluded {\em a priori}. But this raises the problem of why the
Universe rotates so slowly.  This problem can also be naturally
solved in the framework of the inflationary model.  Ellis and
Olive \cite{ElOl83} and Gr{\o}n and Soleng \cite{GrSo87} pointed
out that if the Universe came into being as a mini-universe of
Planck dimensions and went directly into an inflationary epoch
driven by a scalar field with a flat potential, due to the
non-rotation of the false vacuum and the exponential expansion
during inflation the cosmic vorticity has decayed by a factor of
about $10^{-145}$.  The most important diluting effect of the
order of $10^{-116}$ is due to the relative density of the
rotating fluid compared to the non-rotating decay products of the
false vacuum \cite{GrSo87}.  Inflationary cosmology also ruled
out the possibility that the vorticity of galaxies and stars be
of cosmic origin.

While scalar field-driven inflationary models resolve many
problems of the conventional cosmology, inflationary cosmology is
still facing the initial singularity problem.  To solve it
Gasperini and Veneziano \cite{GaVe93} initiated a program, known
as pre-big bang cosmology which is based on the low energy
effective action resulting from string theory.  At the lowest
order in the string frame the NS-NS sector of the four
dimensional string  effective action is given by
\begin{equation}\label{R1}
\hat S = \int d^4 x \sqrt{-\hat g} e^{-2\phi} \left\{ \hat R
   + \hat \kappa ( \nabla \phi )^2 - \frac1{12} \hat H_{[3]}^2
   - \hat U(\phi) \right\},
\end{equation}
where $H_{\mu\nu\lambda}=\partial_{[\mu}B_{\nu\lambda]}$ is the
antisymmetric tensor field, $\hat \kappa$ is the generalized
dilaton coupling constant ($\hat\kappa=4$ for string theory) and
$\hat U(\phi)$ the dilaton potential.  Under certain
circumstances, this low energy string action possesses a symmetry
property, called scale factor duality, which let us expect that
the present phase of the Universe is preceded by an inflationary
pre-big bang phase. Explicit dual solutions can be constructed
for each Bianchi space-time, except Bianchi class A types VIII
and IX models \cite{DiDe99}.

By means of the conformal rescaling $g_{\mu\nu}=e^{-2\phi} \hat
g_{\mu\nu}$, the action (\ref{R1}) can be transformed to the
so-called Einstein frame as
\begin{equation}\label{R2}
S = \int d^4 x \sqrt{-g} \left\{ R - \kappa ( \nabla \phi )^2
   - \frac1{12} e^{-4\phi} H_{[3]}^2 - U(\phi) \right\},
\end{equation}
where $\kappa=6-\hat \kappa$, $U(\phi)=e^{2\phi} \hat U(\phi)$ and
$H_{[3]}^2$ is the square of the antisymmetric field with respect
to the metric $g_{\mu\nu}$.  The $H$-field satisfies the
integrability condition $\partial_{[\mu} H_{\nu\lambda\alpha]}=0$.
Generically, in these type of models the dynamics of the Universe
is dominated by massless bosonic fields.

Starting from the actions (\ref{R1})-(\ref{R2}) cosmological
models in which the Universe starts out in a cold
dilaton-dominated contracting phase, goes through a bounce and
then emerges as an expanding FRW Universe have been explicitly
constructed (for a recent and extensive review of pre-big bang
cosmology see \cite{LiWaCo99}).  Exact solutions for the G\"odel
metric in string theory for the full $O(\alpha')$ action including
both dilaton and axion fields have been obtained by Barrow and
Dabrowski \cite{BaDa98} who also showed that in low energy
effective string theories G\"odel spacetimes need not to contain
closed timelike curves.  According to their results the axion
cannot be introduced in the Einstein frame but plays a crucial
role in the string frame.

It is the purpose of the present paper to investigate in the
framework of the low energy string effective actions
(\ref{R1})-(\ref{R2}) the rotational perturbations of FRW type
cosmological models in both Einstein and string frames and to
find to what extent the possibility of a rotating and expanding
Universe can be incorporated to these type of models. As a
general result we find that for a pure dilaton and axion fields
filled Universe the rotational perturbations always decay due to
the presence of a dilaton field potential. For a potential-free
dilaton field the long-time behaviour of rotation is governed by
an arbitrary function of time, whose explicit mathematical form
cannot be obtained in the framework of the first order
perturbation theory.

The present paper is organized as follows.  In Section II we
obtain the basic equations describing rotational perturbations of
string cosmologies in a flat FRW background. The evolution of
rotational perturbations in Einstein frames is considered in
Section III.  The rotation in the string frame is analyzed in
Section IV. In Section V we conclude our results.

\section{Geometry, Field Equations and Consequences}
In four dimensions, every three-form field can be dualized to a
pseudoscalar.  Thus, an appropriate ans\"atz for the $H$-field is
$H^{\mu\nu\lambda}=\frac1{\sqrt{-g}} e^{4\phi}
\epsilon^{\mu\nu\lambda\rho} \partial_{\rho} h$, where
$\epsilon^{\mu\nu\lambda\rho}=-\delta_{[0}^\mu \delta_1^\nu
\delta_2^\lambda \delta_{3]}^\rho$ and $h=h(t)$ is the Kalb-Ramond
axion field.  The gravitational field equations derived from the
action (\ref{R2}) are
\begin{eqnarray}
R_{\mu\nu} - \kappa \partial_\mu \phi \partial_\nu \phi
   - \frac12 e^{4\phi} \partial_\mu h \partial_\nu h
   - \frac12 g_{\mu\nu} U(\phi) &=& 0, \label{R3} \\
\nabla^2 \phi - \frac1{\kappa} e^{4\phi} (\partial h)^2
   - \frac1{2\kappa} \partial_\phi U(\phi) &=& 0, \label{R4} \\
\partial_\mu \left( \sqrt{-g} e^{4\phi} \partial^\mu h \right)
  &=& 0. \label{R5}
\end{eqnarray}

In the Einstein frame the rotationally perturbed metric can be
expressed in terms of the usual coordinates in the form
\cite{MuFeBr92}
\begin{equation}\label{R6}
ds^2 = - dt^2 + a^2(t) \left[ \frac{dr^2}{1 - k r^2} + r^2 \left(
    d \theta^2 + \sin^2\theta d\varphi^2 \right) \right]
   - 2 \Omega(t,r) a^2(t) r^2 \sin^2\theta \, dt d\varphi,
\end{equation}
where $\Omega(t,r)$ is the metric rotation function, which is
related to local dragging of inertial frames.  We assume that
rotation is sufficiently slow so that deviations from spherical
symmetry can be neglected.  For the sake of mathematical
simplicity and physical clarity we consider only the case of the
flat geometry corresponding to $k=0$.  Then to first order in
$\Omega$ the gravitational and field equations become
\begin{eqnarray}
6\frac{\ddot a}{a} + 2\kappa \dot \phi^2 + e^{4\phi} \dot h^2
   - U(\phi) &=& 0, \label{R7} \\
2\frac{\ddot a}{a} + 4\frac{\dot a^2}{a^2}
   - U(\phi) &=& 0, \label{R8} \\
\ddot \phi + 3\frac{\dot a \dot \phi}{a}
   - \frac1{\kappa} e^{4\phi} \dot h^2
   + \frac1{2\kappa} \partial_\phi U(\phi) &=& 0, \label{R9} \\
\ddot h + 4 \dot h \dot \phi
   + 3\frac{\dot a \dot h}{a} &=& 0, \label{R10} \\
3\frac{\dot a}{a} \partial_r \Omega(t,r)
   + \partial_t \partial_r \Omega(t,r) &=& 0, \label{R11} \\
\partial_r^2 \Omega(t,r) + \frac4{r} \partial_r \Omega(t,r)
   - \Omega(t,r) \left(2\ddot a a+4\dot a^2\right) &=& 0. \label{R12}
\end{eqnarray}

The axion and dilaton field equations are unperturbed to the first
order in $\Omega$.  This justifies the assumption of homogeneity
in the rotation to first order.  Eqs. (\ref{R11}) and (\ref{R12})
follow from the \{13\} and \{03\} components of the Einstein
equation (\ref{R3}) respectively.

Eq. (\ref{R10}) can be integrated to give
\begin{equation}\label{R13}
\dot h = C e^{-4\phi} a^{-3},
\end{equation}
with $C$ an arbitrary constant of integration.  Using Eq.
(\ref{R13}) the evolution equation of the dilaton field
(\ref{R9}) becomes
\begin{equation}\label{R14}
2\kappa a^{-3} \partial_t \left( a^3 \dot \phi \right)
   - 2 C^2 e^{-4\phi} a^{-6} + \partial_\phi U(\phi) = 0.
\end{equation}
Multiplication of Eq. (\ref{R14}) with $a^6 \dot \phi$ leads to
the following first integral of the dilaton field equation:
\begin{equation}\label{a1}
2 \kappa a^6 \dot \phi^2 + C^2 e^{-4\phi}
   + 2 \int a^6 \dot U(t) dt = 2 \kappa \phi_0,
\end{equation}
with $\phi_0$ a constant of integration. The elimination of the
term $6\ddot a/a$ between the field equations (\ref{R7}) and
(\ref{R8}) gives
\begin{equation}\label{a2}
12 a^4 \dot a^2 - 2 \kappa a^6 \dot\phi^2 - C^2 e^{-4\phi}
   - 2 a^6 U(\phi) = 0.
\end{equation}

Therefore from Eqs. (\ref{a1}) and (\ref{a2}) we obtain the
following consistency condition relating the scale factor $a(t)$
to the dilaton field potential
\begin{equation}\label{a3}
6 a^4 \dot a^2 + \int a^6 \dot U(t) dt
   = a^6 U(\phi) + \kappa \phi_0.
\end{equation}

In the present paper we are primarily interested in the evolution
of the metric rotation function $\Omega(t,r)$.  Eq. (\ref{R11})
can be integrated to obtain
\begin{equation}\label{R15}
\Omega(t,r) = F(r) a^{-3}(t) + G(t),
\end{equation}
with $G(t)$ an arbitrary function of time.  Therefore from Eqs.
(\ref{R12}) and (\ref{R8}) it follows that the function $F(r)$
must obey the equation
\begin{equation}\label{R16}
\partial_r^2 F(r) + \frac4{r} \partial_r F(r)
   - a^2(t) U(\phi) \left[ F(r) + G(t) a^3(t) \right] = 0.
\end{equation}

\section{Evolution of Rotational Perturbations in the Einstein Frame}
The behavior of the rotational perturbations of the isotropic flat
FRW cosmological models essentially depends upon the dilaton
potential.  The form (\ref{R16}) of the equation governing the
spatial part of the metric rotation function imposes strong
constraint on the functional form of $U(\phi)$. There are {\em
two and only two} forms of the dilaton field potential which make
the rotation equation (\ref{R16}) mathematically consistent.

\subsection{Case I: $U(\phi)=0$}
For $U(\phi) \equiv 0$, Eq. (\ref{R16}) can be immediately
integrated to give
\begin{equation}\label{R17}
F(r) = F_1 - F_0 r^{-3},
\end{equation}
with $F_0, F_1$ arbitrary constants of integration.  From the
field equation (\ref{R8}) we obtain the scale factor $a(t)$ in
the form
\begin{equation}\label{R18}
a(t) = a_0 t^{1/3},
\end{equation}
where $a_0$ is an arbitrary constant of integration.  For this
model the deceleration parameter $q$, defined as $q=d
H^{-1}/dt-1$, $H=\dot a/a$, is given by $q=2$.  The sign of the
deceleration parameter shows that the cosmological model inflates
or not --- negative sign for the inflationary models while
positive sign corresponding to the standard decelerating models.
Therefore in the absence of a dilaton potential the cosmological
evolution of the axion and dilaton fields filled slowly rotating
Universe is non-inflationary.

With a vanishing dilaton potential the first integral (\ref{a1})
of the dilaton field equation leads to the following general
representation of the conformal transformation factor:
\begin{equation}\label{R19}
e^{2\phi(t)} = \sqrt{\frac{C^2}{2\kappa\phi_0}}
    \cosh \left( 2\sqrt{\phi_0} \int \frac{dt}{a^3} \right),
\end{equation}
With the use of (\ref{R18}) we find
\begin{equation}\label{R20}
e^{2\phi(t)} = \sqrt{\frac{C^2}{8\kappa\phi_0}}
    \left( t^\sigma + t^{-\sigma} \right),
\end{equation}
where $\sigma=2\sqrt{\phi_0} a_0^{-3}$. For the Kalb-Ramond field
we obtain
\begin{equation}\label{R21}
h(t) = h_0 - \frac{2\kappa\sqrt{\phi_0}}{C}
    \frac1{t^{2\sigma}+1}.
\end{equation}
Eq. (\ref{a3}) gives the following consistency condition for the
integration constants:
\begin{equation}\label{b1}
a_0^6=\frac32 \kappa \phi_0.
\end{equation}

The metric rotation function behaves like
\begin{equation}\label{R22}
\Omega(t,r) = a_0^{-3} t^{-1} \left( F_1 - F_0 r^{-3} \right)
  + G(t).
\end{equation}
In the large time limit the behavior of the rotational
perturbations of the axion and dilaton fields filled Universe is
governed in the Einstein frame by the arbitrary function $G(t)$,
$\lim_{t\to\infty} \Omega(t,r)=\lim_{t\to\infty}G(t)$. In order
to fix our original assumption of small rotation, we would expect
that the function $G(t)$ has the faster or at least equal fall off
$1/t$.

\subsection{Case II: $U(\phi)=U_0^2 a^{-2}(t)$}
The second case in which the rotation equation (\ref{R16}) can be
integrated is for a dilaton field potential satisfying the
condition
\begin{equation}\label{R23}
U(\phi) = U_0^2 a^{-2}(t),
\end{equation}
with constant $U_0$. Then Eq. (\ref{R16}) fixes the arbitrary
function $G(t)$ as
\begin{equation}
G(t) U(\phi) a^5(t) = U_0^2 G(t) a^3(t) = G_0 U_0^2,
\end{equation}
with $G_0$ an arbitrary constant. Consequently $G(t)=G_0
a^{-3}(t)$. With this choice Eq. (\ref{R16}) becomes
\begin{equation}\label{R24}
\partial_r^2 F(r)+\frac4{r} \partial_r F(r)-U_0^2 F(r) = G_0 U_0^2.
\end{equation}
In order to solve Eq. (\ref{R24}) we introduce a new function
$Y(r)$ by means of the transformation $Y(r)=F(r)+G_0$. Then
$Y(r)$ satisfies the differential equation
\begin{equation}\label{EY}
\partial_r^2 Y(r) + \frac4{r} \partial_r Y(r) - U_0^2 Y(r) = 0.
\end{equation}

Let $f(r)$ be the general solution of the equation $\partial_r^2
f(r) - U_0^2 f(r)=0$. Then we can represent the general solution
of Eq. (\ref{EY}) in the form $Y(r)=C_1 r^{-2} f(r) - C_2 r^{-3}
\partial_r f(r)$, with two constants $C_1, C_2$.  Substitution
into Eq. (\ref{EY}) leads to a possible choice $C_1=U_0^2,
C_2=1$.  Therefore the spatial part of the rotation function is
given by
\begin{equation}\label{R25}
F(r) = r^{-3} \left[ U^2_0 r \left( Ae^{U_0 r} + Be^{-U_0 r}
  \right) - U_0 \left( Ae^{U_0 r} - Be^{-U_0 r} \right) \right]
  - G_0,
\end{equation}
where $A$ and $B$ are arbitrary constants of integration.  This
solution is not regular in the origin $r=0$.

The evolution of the scale factor of the Universe can be obtained
from Eq. (\ref{R8}), which with the potential (\ref{R23}) becomes
\begin{equation}\label{R26}
2 a \ddot a + 4 \dot a^2 = U_0^2,
\end{equation}
and has the general solution
\begin{equation}\label{R27}
t - t_0 = \int \frac{a^2 da}{\sqrt{K+U_0^2 a^4/4}},
\end{equation}
with $K$ an arbitrary constant of integration. Substitution in
Eq. (\ref{a3}) gives the consistency condition
\begin{equation}\label{b2}
6K = \kappa \phi_0.
\end{equation}

The solution of Eq. (\ref{R27}) can be represented in terms of
elliptical functions, but in order to have a better physical
insight we consider only the solution corresponding to the large
time behaviour of the model, when the condition $U_0^2 a^4/4>>K$
holds with a very good approximation.  This is equivalent to
taking the arbitrary integration constant $K=0$.  Consequently
from (\ref{b2}) we also have $\phi_0=0$.  Therefore in this limit
the Einstein frame time evolution of the scale factor of the
rotationally perturbed flat FRW model is given by
\begin{equation}\label{R28}
a(t) = \frac{U_0}2 t.
\end{equation}
In this case the deceleration parameter is given by $q=0$.  The
evolution of the Universe is at the exact limit separating
inflationary and non-inflationary phases.

The dilaton field equation (\ref{a1}) can be written in the form
\begin{equation}\label{R29}
\kappa U_0^6 t^6 \dot \phi^2 + 32 C^2 e^{-4\phi}
  - 2 U_0^6 t^4 = 0.
\end{equation}
By introducing a new variable $y=e^{-\phi}$, Eq.(\ref{R29})
becomes
\begin{equation}\label{dy}
\dot y^2 + \frac{32C^2}{\kappa U_0^6} \left(\frac{y}{t}\right)^6
  - \frac2{\kappa} \left(\frac{y}{t}\right)^2 = 0.
\end{equation}
With the substitution $y=ut$ and by denoting
$\Delta(u)=\sqrt{1-16C^2 u^4/U_0^6}$, Eq. (\ref{dy}) is
transformed into
\begin{equation}
t\frac{du}{dt}=\sqrt{2/\kappa} \, u(\Delta(u)-\sqrt{\kappa/2}),
\end{equation}
with the general solution given by
\begin{eqnarray}
t &=& t_0
\frac{(\Delta(u)-1)^\alpha}{(\Delta(u)-\sqrt{\kappa/2})^\beta
      (\Delta(u)+1)^\gamma}, \qquad  \kappa \neq 2, \\
t &=& t_0 \exp \left[ -\frac1{4(\Delta(u)-1)} \right]
      \left( \frac{\Delta(u)-1}{\Delta(u)+1} \right)^{1/8},
      \qquad \kappa=2,
\end{eqnarray}
where $\alpha^{-1}=4(\sqrt{2/\kappa}-1),
\beta^{-1}=2(2/\kappa-1), \gamma^{-1}=4(\sqrt{2/\kappa}+1)$ and
$t_0$ is a constant of integration.

Therefore the evolution of the dilaton field, dilaton potential
and axion field can be represented in the following exact
parametric form:
\begin{equation}
e^{2\phi}=u^{-2} t^{-2}, \qquad U(\phi)=4 t^{-2},
\end{equation}
and
\begin{eqnarray}
h(u) &=& \sqrt{\frac{\kappa}2}\frac{8Ct_0^2}{U_0^3}
    \int \frac{u^3 (\Delta(u)-1)^{2\alpha} \, du}
         {(\Delta(u)-\sqrt{\kappa/2})^{2\beta+1}
          (\Delta(u)+1)^{2\gamma}}, \qquad \kappa\neq 2, \\
h(u) &=& \sqrt{\frac{\kappa}2}\frac{8Ct_0^2}{U_0^3}
    \int u^3 \exp \left[ -\frac1{2(\Delta(u)-1)} \right]
     \frac{\left(\Delta(u)-1\right)^{-3/4}}
          {\left(\Delta(u)+1\right)^{1/4}} \, du, \qquad \kappa=2,
\end{eqnarray}
respectively.  The dilaton and the axion fields are defined only
for values of the parameter $u$ so that $u\leq\frac12
U_0^{3/2}C^{-1/2}$. Therefore during the cosmological evolution
the dilaton field satisfies the condition $e^\phi\geq
2C^{1/2}U_0^{-3/2}t^{-1}$.  For the dilaton potential we obtain
$U(\phi)\leq U_0^3C^{-1}e^{2\phi}$.

The time evolution of the dilaton and Kalb-Ramond axion fields are
represented in Figs. 1 and 2, respectively.  The dynamics of these
fields essentially depend on the string coupling constant
$\kappa$. For $\kappa<2$ the dilaton field tends to infinity in
the small time limit and for large $t$ decreases rapidly to
zero.  For $\kappa\ge 2$, however, the dilaton field is zero at
$t=t_0=0$ and is a monotonically increasing function of time. The
time variation of the axion shows an opposite dynamics.  For
$\kappa<2$ the axion field is zero at the initial stages of the
evolution of the Universe and then it rapidly increases in time.
For $\kappa\ge 2$ the axion field tends to infinity for $t=t_0=0$
but in the large time limit $h\to 0$.

The dilaton field potential can be expressed as a function of the
dilaton field $\phi$ in the form $U(\phi)=4u^2 e^{2\phi}$.
Generally the dilaton field potential can be represented in the
form $U(\phi)=g(\phi) e^{2\phi}$, with $g(\phi)$ a function which
does not have, in the present case, an analytical
representation.  For intervals of time when $u$ can be considered
a constant or a slowly varying function of time, the potential -
dilaton field dependence is of pure exponential type.

The rotation function is
\begin{equation}\label{R32}
\Omega(t,r) = 8 U_0^{-3} t^{-3} r^{-3}
     \left[ U^2_0 r \left( Ae^{U_0 r} + Be^{-U_0 r} \right)
   - U_0 \left( Ae^{U_0 r} - Be^{-U_0 r} \right) \right].
\end{equation}

The time decay of the rotation is inverse proportional to the
third power of the time.  In the large time limit we have
$\lim_{t\to\infty} \Omega(t,r)=0$.  Therefore in the Einstein
frame in the presence of an exponential type dilaton field
potential there is a rapid decay of the rotational perturbations
of the FRW Universe.

For $C=0$, that is, for a constant axion field, which can be
chosen, without any loss of generality, to be zero, Eq.
(\ref{R29}) gives
\begin{equation}\label{AA}
e^{2\phi(t)} = \varphi_0^2 t^{\sqrt{8/\kappa}},
\end{equation}
$\varphi_0=\hbox{const.}$ In this case in the large time limit
the dilaton field is an increasing function of time.

It is interesting to note that if the dilaton potential is a
non-zero constant, $U(\phi) \equiv \Lambda$, then the rotation
equation (\ref{R16}) implies $a=\hbox{const.}$  In this case the
field equations become inconsistent unless $\Lambda=0$. Therefore
the first order rotational perturbations of the FRW cosmological
models does not support the existence of a cosmological constant
in the Einstein frame.

\begin{figure}
\epsfxsize=9cm \centerline{\epsffile{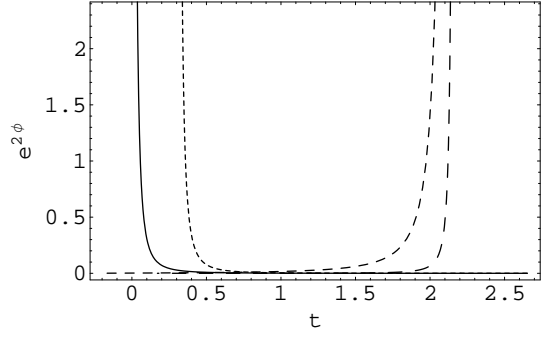}} \caption{Einstein
frame time evolution of the conformal transformation factor
$e^{2\phi}$ (in units of $10^6$) in the case of non-vanishing
potential dilaton field for different values of the parameter
$\kappa$: $\kappa=1$ (solid curve), $\kappa=3/2$ (dotted curve),
$\kappa=2$ (short dashed curve) and $\kappa=3$ (long dashed
curve). We have used the normalizations $16C^2/U_0^6=1$ and
$t_0=1$.} \label{FIG1}
\end{figure}

\begin{figure}
\epsfxsize=9cm \centerline{\epsffile{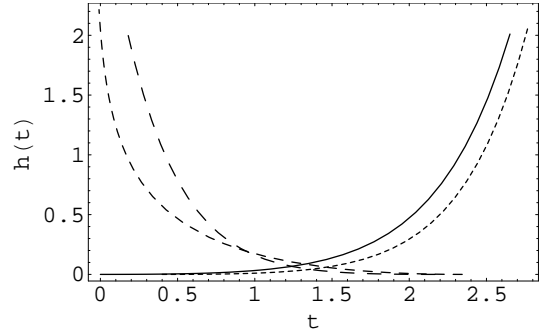}} \caption{Einstein
frame dynamics of the axion field $h$ (in units of $10^6$) for
the non-vanishing dilaton potential for different values of the
string coupling constant $\kappa$: $\kappa=1$ (solid curve),
$\kappa=3/2$ (dotted curve), $\kappa=2$ (short dashed curve) and
$\kappa=3$ (long dashed curve). We have used the normalizations
$16C^2/U_0^6=1$ and $t_0=1/\sqrt{2}$.} \label{FIG2}
\end{figure}

\section{Rotational Perturbations in the String Frame}
We consider now the evolution of the rotational metric function
in the string frame.  In this frame the components of the metric
tensor are given by $\hat g_{\mu\nu}=e^{2\phi(t)}g_{\mu\nu}$.  If
we define a new time variable $\hat t$ by means of the
transformation $d\hat t=e^{\phi(t)}dt$, then in the string frame
the line element of the rotationally perturbed metric is given by
\begin{equation}\label{R33}
d \hat s^2 = - d \hat t^2 + \hat a^2(\hat t) \left[
  \frac{dr^2}{1-kr^2}
    + r^2 \left( d\theta^2 + \sin^2\theta d\varphi^2 \right) \right]
    - 2 \Omega(\hat t,r) \hat a^2(\hat t) e^{-\phi(\hat t)}
      r^2 \sin^2\theta \, d\hat t d\varphi,
\end{equation}
where $\hat a^2(\hat t)=e^{2\phi(\hat t)} a^2(\hat t)$.

In the case of the potential free dilaton field the string frame
cosmological time $\hat t$ is defined according to $\hat
t=\left(\frac{C^2}{8\kappa\phi_0}\right)^{1/4} \int
\sqrt{t^{\sigma}+t^{-\sigma}} dt$. In the large time limit we
obtain $\hat t \sim \hat t_0 t^{(\sigma+2)/2}$, with $\hat
t_0=\frac2{\sigma+2} \left(\frac{C^2}{8\kappa\phi_0}
\right)^{1/4}$. The cosmological time in the string frame is
proportional to the cosmological time in the Einstein frame.  The
conformal transformation factor, also describing the dilaton field
evolution, becomes $e^{\phi} \sim
\left(\frac{C^2}{8\kappa\phi_0}\right)^{1/4} \left( \frac{\hat t}
{\hat t_0} \right)^{\sigma/(\sigma+2)}.$ For the string frame
scale factor we find
\begin{equation}\label{R34}
\hat a(\hat t) \sim \hat a_0 \left( \frac{\hat t}{\hat t_0}
  \right)^{(3\sigma+2)/(3\sigma+6)},
\end{equation}
where $\hat a_0=a_0
\left(\frac{C^2}{8\kappa\phi_0}\right)^{\frac14}$. The metric
tensor component $\hat g_{03}$ can be represented as
\begin{equation}
\hat g_{03}
  \sim -2\left(\frac{C^2}{8\kappa\phi_0}\right)^{-1/4}
  \left[a_0^{-1}\left(\frac{\hat t}{\hat t_0}\right)^{\frac{3\sigma-2}
    {3\sigma+6}}\left(F_1-F_0 r^{-3}\right)+a_0^2\left(
    \frac{\hat t}{\hat t_0}\right)^{\frac{3\sigma+4}{3\sigma+6}}
    \hat G\left(\hat t\right) \right] \, r^2\sin^2\theta.
\end{equation}

The string frame decay of the rotational perturbations is again
governed by the arbitrary function $\hat G(\hat t)$.  The
Kalb-Ramond field is $h(\hat t)\sim h_0$ approaching to a
constant.

The second situation for which the rotation equation describing
the string frame evolution of the slowly rotating Universe has a
solution corresponds to the exponential type dilaton potential. We
restrict again our analysis to the $K=0$ case only.  Then
defining the cosmological string frame time as $\hat t=\hat t_0 +
\int e^\phi dt, \, \hat t_0=\hbox{const.}$ and introducing a new
variable $\eta=\sqrt{2/\kappa}\Delta(u)$, $0 \leq \eta \leq
\sqrt{2/\kappa}$, the general solution of the field equations can
be represented in the following exact parametric form:
\begin{eqnarray}
\hat t(\eta) &=& \hat t_0 + \frac{\kappa\sqrt{C}}{2U_0^{3/2}} \int
  \frac{\eta d\eta}{(1-\eta)\left(1-\kappa\eta^2/2\right)^{5/4}},
  \label{hatt} \\
\hat a(\eta) &=& \sqrt{C/U_0}\left(1-\kappa\eta^2/2\right)^{-1/4},\\
e^{\phi(\eta)} &=& \frac{2\sqrt{C}}{t_0 U_0^{3/2}}
  \left(1-\kappa\eta^2/2\right)^{-1/4} \exp \left[ -\frac{\kappa}4
  \int \frac{\eta d\eta}{(1-\eta)\left(1-\kappa\eta^2/2\right)}
  \right], \\
\hat U(\eta) &=& \frac{U_0^3}{C}
  \left(1-\kappa\eta^2/2\right)^{1/2}, \\
h(\eta) &=& \frac{\kappa t_0^2 U_0^3}{8C} \int
  \frac{\eta \exp\left[\frac{\kappa}2 \int \frac{\eta d\eta}{(1-\eta)
  \left(1-\kappa\eta^2/2\right)}\right]}{1-\eta} d\eta.
\end{eqnarray}

The spatial distribution of the rotation function is described by
Eq. (\ref{R25}).  The temporal behavior of the string frame
metric tensor component $\hat g_{03}$ is governed by the function
\begin{equation}
\hat f(\eta) \sim \left(1-\kappa\eta^2/2\right)^{-1/4}
  \exp \left[ -\frac{\kappa}2 \int \frac{\eta d\eta}{(1-\eta)
  \left(1-\kappa\eta^2/2\right)}\right].
\end{equation}

In the limit of small $\eta$, $\eta\to 0 \, (u\to 1)$, from Eq.
(\ref{hatt}) we obtain $\kappa\eta^2/2\approx 2U_0^{3/2} C^{-1/2}
\hat t$.  Therefore in the small time limit the scale factor
behaves like $\hat a(\hat t) \approx \sqrt{C/U_0}
\left(1-2U_0^{3/2}C^{-1/2} \hat t\right)^{-1/4}$. In the string
frame and in the presence of the exponential type dilaton
potential the slowly rotating Universe starts its evolution from
a non-singular state with $\hat a(0)=\sqrt{C/U_0} \neq 0$.  For
the dilaton and axion fields we obtain $e^{\phi(\hat t)}\approx
2\sqrt{C/U_0^3}\exp\left(-\sqrt{U_0^3/C}\hat t/2\right)$ and
$h(\hat t)\approx \hat h_0+(U_0^6 t_0^2/32C)
\exp\left(\sqrt{U_0^3/C}\hat t\right)$, respectively.  At the
early beginning of the Universe the time behavior of $\hat
g_{03}$ is given by $\hat f(\hat t) \approx
\exp\left(-2\sqrt{U_0^3/C} \hat t\right)$.

The time evolutions of the scale factor $\hat a$, dilaton field
$\phi$, dilaton potential $\hat U$, axion field $h$ and $\hat
g_{03}$ as functions of the string frame cosmological time are
represented in Figs.  3-7.  The scale factor is represented in
Fig. 3 for different $\kappa$.  In the string frame the evolution
of the universe starts from a nonsingular state, with the
Einstein frame singularity of the scale factor $a(t)=U_0t/2$
removed by the conformal transformation $e^{\phi}a=1/u$.  The
dynamics of the dilaton field potential, presented in Fig.  4,
shows a monotonically decrease of $\hat U$ and in the large time
limit $\hat U\to 0$.  In Fig. 5 we represented the dynamics of
the dilaton field $\phi$.  Independently of the values of the
string coupling constant $\kappa$, in the large time limit we have
$\phi \to 0$.  Hence at the end of the cosmological evolution
both the dilaton field and dilaton field potential vanishes.  But
the axion field time variation, presented in Fig. 6, shows a
rapid time increase of $h$, with $h\to\infty$ in the large time
limit.  Therefore in this model in the large time limit the
dynamics of the Universe is governed by the Kalb-Ramond axion
field.

In the string frame the time behavior of the rotational
perturbations, presented in Fig.  7 is similar to that in the
Einstein frame, in the large time limit $\hat g_{03}\to 0$.  If
in the Einstein frame the time decay of the rotation is given by
a power law, being proportional to $t^{-3}$, in the string frame
the first order rotational perturbations decay exponentially.

The string frame deceleration parameter $\hat q$ is given, as a
function of $\eta$, by
\begin{equation}
\hat q(\eta) = \frac4{\kappa} \frac{1-\kappa\eta^2/2}
  {\eta (1-\eta)}.
\end{equation}
The time evolution of the deceleration parameter is represented
in Fig. 8. The dynamics of $\hat q$ is dependent of the string
coupling constant $\kappa$.  For $\kappa<2$, since values of
$\eta>1$ are allowed the universe ends in the large time limit in
an inflationary phase.  For $\kappa\ge 2$ generally $\hat q>0$
and hence the string frame evolution is non-inflationary.  In
this case in the large time limit $\hat q \to 0$ and therefore
the Universe ends at the exact limit separating inflationary and
non-inflationary phases.

For a vanishing axion field the general solution in the string
frame can be obtained in an exact form.  The string frame
cosmological time $\hat t$ is related to the Einstein frame
cosmological time $t$ by means of the relation $\hat
t=\frac{\varphi_0}{\sqrt{2/\kappa}+1}\,t^{\sqrt{2/\kappa}+1}$.
The scale factor has the same behavior as in the Einstein frame,
being given by
\begin{equation}
\hat a(\hat t) = \hat a_0 \hat t,
\end{equation}
with $\hat a_0=U_0(\sqrt{2/\kappa}+1)/2$. For this model in the
string frame the rotational perturbations
\begin{equation}
\hat g_{03} \sim \hat
  t^{\frac{\sqrt{2/\kappa}-1}{\sqrt{2/\kappa}+1}}
   \left[ U_0^2 r \left( A e^{U_0 r} + B e^{-U_0 r} \right)
   - U_0 \left( A e^{U_0 r} - B e^{-U_0 r} \right) \right]
   r^{-1} \sin^2\theta,
\end{equation}
may decay ($\kappa>2$), increase ($\kappa<2$) or be independent
($\kappa=2$) with respect to time $\hat t$.

\begin{figure}
\epsfxsize=9cm \centerline{\epsffile{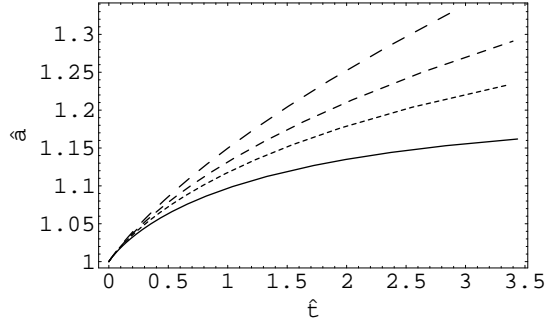}} \caption{String
frame time evolution of the scale factor $\hat a$ in the case of
non-vanishing potential dilaton field for different values of the
parameter $\kappa$: $\kappa=1$ (solid curve), $\kappa=3/2$ (dotted
curve), $\kappa=2$ (short dashed curve) and $\kappa=3$ (long
dashed curve). We have used the normalizations
$\frac{\sqrt{C}}{2U_{0}^{3/2}}=1$ and $\sqrt{C/U_{0}}=1$, leading
to $C=U_0=1/2$.} \label{FIG3}
\end{figure}

\begin{figure}
\epsfxsize=9cm \centerline{\epsffile{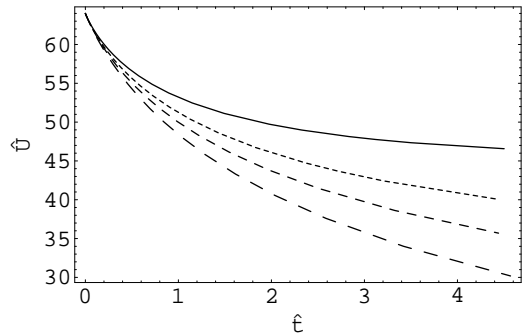}} \caption{String
frame time dynamics of the dilaton field potential $\hat U$ for
different values of the parameter $\kappa$: $\kappa=1$ (solid
curve), $\kappa=3/2$ (dotted curve), $\kappa=2$ (short dashed
curve) and $\kappa=3$ (long dashed curve). We have used the
normalization $U_0^3/C=1/4$.} \label{FIG4}
\end{figure}

\begin{figure}
\epsfxsize=9cm \centerline{\epsffile{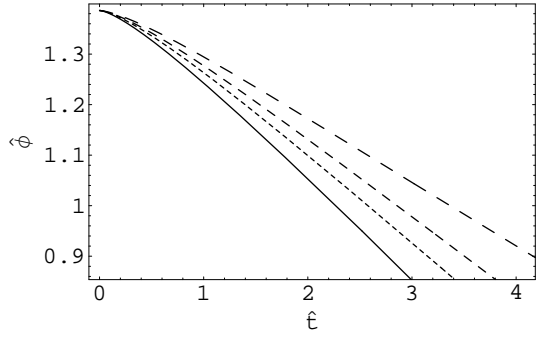}}
\caption{Variation of the dilaton field $\phi$ as a function of
the string frame cosmological time $\hat t$ in the case of
non-vanishing dilaton potential for different values of the
parameter $\kappa$: $\kappa=1$ (solid curve), $\kappa=3/2$
(dotted curve), $\kappa=2$ (short dashed curve) and $\kappa=3$
(long dashed curve). We have used the normalizations $t_0=1$ and
$\frac{\sqrt{C}}{2U_{0}^{3/2}}=1$.} \label{FIG5}
\end{figure}

\begin{figure}
\epsfxsize=9cm \centerline{\epsffile{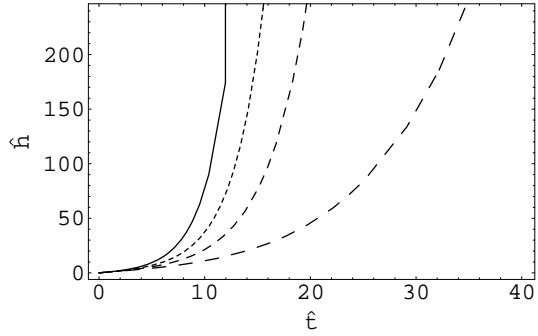}} \caption{Time
evolution of the axion field $h$ as a function of string frame
cosmological time in the case of non-vanishing potential dilaton
field for different values of the parameter $\kappa$: $\kappa=1$
(solid curve), $\kappa=3/2$ (dotted curve), $\kappa=2$ (short
dashed curve) and $\kappa=3$ (long dashed curve). We have used
the normalizations $t_0=4\sqrt{2}$ and
$\frac{\sqrt{C}}{2U_{0}^{3/2}}=1$.} \label{FIG6}
\end{figure}

\begin{figure}
\epsfxsize=9cm \centerline{\epsffile{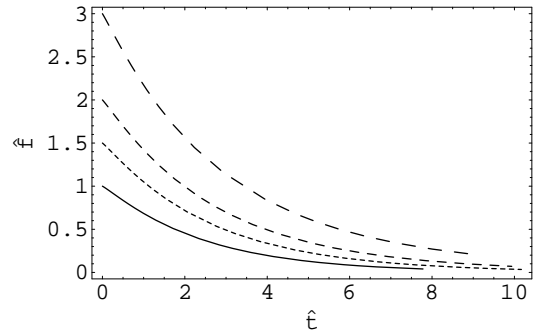}} \caption{Time
decay of the rotational perturbations $\hat f$ as a function of
the string frame cosmological time in the case of non-vanishing
potential dilaton field for different values of the parameter
$\kappa$: $\kappa=1$ (solid curve), $\kappa=3/2$ (dotted curve),
$\kappa=2$ (short dashed curve) and $\kappa=3$ (long dashed
curve). We have used the normalization
$\frac{\sqrt{C}}{2U_{0}^{3/2}}=1$.} \label{FIG7}
\end{figure}

\begin{figure}
\epsfxsize=9cm \centerline{\epsffile{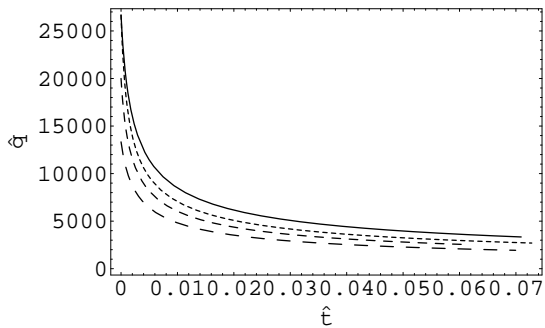}} \caption{String
frame evolution of the deceleration parameter $\hat q$ for
different values of the string coupling constant $\kappa$:
$\kappa=1$ (solid curve), $\kappa=3/2$ (dotted curve), $\kappa=2$
(short dashed curve) and $\kappa=3$ (long dashed curve). We have
used the normalization $\frac{\sqrt{C}}{2U_{0}^{3/2}}=1$.}
\label{FIG8}
\end{figure}

\section{Discussions and Final Remarks}
In this paper we have analyzed rotational perturbations of
homogeneous flat isotropic FRW in NS-NS string cosmology. To
first order in the metric rotation function the field equations
reduce to the unperturbed field equations in addition to two
equations determining the rotation function $\Omega(t,r)$. The
decay of the rotation is basically determined by the dilaton
field potential. The field equations impose strong constraint
upon the functional form of $U(\phi)$, and except the trivial
form $U(\phi)=0$ only one other form is allowed by the
mathematical structure of the theory. In the Einstein frame the
rotation function can be generally represented as a product of two
independent functions, one depending on time and the other on
$r$, plus a function depending on the cosmological time only.

If the dilaton field potential is zero, the large-time evolution
of the rotational perturbations is determined by an arbitrary
function of time $G(t)$ in both Einstein and string frames.
Therefore in this case the initial rotation of the Universe may
not decay to zero in the large time limit and thus the
possibility of a global rotation in the present day Universe is
not excluded in this model. On the other hand the arbitrary
character of the function $G(t)$ and the absence of a physical
mechanism excluding in a natural way rapid late-time rotation of
the Universe raises the question if a dilaton potential free
pre-big bang type cosmological model can lead to a correct
description of the dynamics of our Universe. This situation is
somehow similar to the behaviour of the anisotropy in the Bianchi
type I space-times in pre-big bang cosmological models. In the
absence of the dilaton field potential a Bianchi type I
space-time does not isotropize, the geometry being of Kasner type
for all times \cite{CHM00a,CHM00b}. Therefore in standard pre-big
bang cosmological models with pure dilaton and axion fields nor
the initial rotation neither the initial anisotropies can be
washed out as a result of the expansionary evolution of the
Universe.

For a non-zero dilaton potential the rotation Eq. (\ref{R16}) and
the field equations completely determine the form of the
potential. Therefore $U(\phi)$ is not an arbitrary parameter of
the theory. In the Einstein frame the dilaton field potential,
can be represented, as a function of time, in a parametric form,
with $a$ taken as parameter, as
\begin{equation}
t-t_0 = \int \frac{a^2 da}{\sqrt{K+U_0^2 a^4/4}}, \qquad
  U(a)=U_0^2 a^{-2}.
\end{equation}
This mathematical form of the potential is the only one allowed
by the mathematical structure of the field equations. In the
limit of small $a$, $a\to 0$, the time dependence of the potential
is given by $U(t)=U_0^2(3\sqrt{K}\,t)^{-2/3}$. In the limit of
large $t$, corresponding to large $a$, we obtain $U(t)=4/t^2$.
The behaviour of the dilaton field $\phi$ is described, as a
function of the parameter $a$, by the equation
\begin{equation}
2\kappa a^2 \left(K+\frac{U_0^2}4 a^4\right)
  \left( \frac{d\phi}{da} \right)^2 + C^2 e^{-4\phi}-U_0^2 a^4
  = 2\kappa \phi_0,
\end{equation}
which follows from Eq. (\ref{a1}). The potential cannot be
expressed as a function of the dilaton field in terms of
elementary functions. In the small-time limit $a\to 0$ the dilaton
field obeys the equation
\begin{equation}
2\kappa K a^2\left(\frac{d\phi}{da}\right)^2=2\kappa \phi_0
  - C^2 e^{-4\phi},
\end{equation}
with the general solution
\begin{equation}
e^{2\phi}=\sqrt{\frac{C^2}{8\kappa \phi_0}} \left(
 a^{2\sqrt{\phi_0/K}} + a^{-2\sqrt{\phi_0/K}}\right) \approx
 \sqrt{\frac{C^2}{8\kappa \phi_0}} a^{-2\sqrt{\phi_0/K}}.
\end{equation}
Therefore in the small-time limit the time dependence of the
conformal transformation factor in the Einstein frame is
\begin{equation}
e^{2\phi} \approx \sqrt{\frac{C^2}{8\kappa \phi_0}}
 \left(3\sqrt{K}\,t\right)^{-\frac23 \sqrt{\phi_0/K}},
\end{equation}
leading to a potential-dilaton field dependence of the
exponential form
\begin{equation}
U(\phi) \approx U_0^2 \left(\frac{8\kappa \phi_0}{C^2}
  \right)^{\frac12\sqrt{K/\phi_0}} e^{2\sqrt{K/\phi_0}
  \, \phi} = \Lambda e^{2\chi \phi},
\end{equation}
where $\Lambda=U_0^2 \left(\frac{8\kappa \phi_0}{C^2}
\right)^{\frac12\sqrt{K/\phi_0}}=\hbox{const.}$ and
$\chi=\sqrt{K/\phi_0}=\hbox{const.}$.  The exponential type
potentials play an important role in particle physics and
cosmology. An exponential potential arises in the
four-dimensional effective Kaluza-Klein theories from
compactification of the higher-dimensional supergravity or
superstring theories. In string or Kaluza-Klein theories the
moduli fields associated with the geometry of the
extra-dimensions may have effective exponential potentials due to
the curvature of the internal spaces or to the interaction of the
modului with form fields on the internal spaces. Exponential
potentials can also arise due to the non-perturbative effects
such as gaugino condensation (for a discussion on the role of
exponential potentials in cosmology see \cite{CHM00a} and
references therein). In the large time limit the potential is a
generalized exponential one, of the form $U(\phi)=g(\phi)
e^{2\phi}$. The function $g(\phi)$ cannot be expressed in terms
of elementary functions. In the string frame and in the same limit
the dilaton field potential can be represented as $\hat U(\phi) =
e^{2\phi}U(\phi)=4e^{2\phi}/t^2$. In the presence of the dilaton
field potential there is a rapid decay of the rotational
perturbations. The rotation equation (\ref{R16}) fixes not only
the form of the potential but also determines the mathematical
form of the function $G(t)$, which in the Einstein frame is
inversely proportional to the third power of the scale factor.
Hence in an expanding Universe the rotational perturbations
decrease rapidly, the long-time decay of the rotation being given
by a power law in the Einstein and by an exponential term in the
string frame.  The general physical requirement of the very small
(or zero) rotation of the late-time Universe also imposes the
presence of the axion field, together with the dilaton field
potential, in the very early stages of cosmological evolution.

If the Kalb-Ramond field $h$ is zero, even in the presence of the
dilaton field potential, the rotational perturbations in the
string frame are governed by the numerical value of the string
coupling constant $\kappa$. In this case the Universe is not
rotating for large cosmological times only if $\kappa>2$. For
string theory $\kappa=2$ and for a zero $h$-field the
corresponding dilatonic Universe rotates for all times.

The observational evidence that our Universe is rotating very
slowly or at all imposes a major constraint on realistic
cosmological models. The first order rotational perturbation
theory analyzed in the present paper could be relevant for the
understanding of the transient period from a rotating initial
state of our Universe to an expansionary one.

\section*{Acknowledgments}
The work of CMC was supported by the Taiwan CosPA project.


\end{document}